\documentclass[prb,twocolumn,showpacs]{revtex4}     
\usepackage{epsfig,amsmath} 

\begin{document}

\title{Current orderings of interacting electrons in bilayer graphene} 

\author{Xin-Zhong Yan$^{1}$ and C. S. Ting$^2$}
\affiliation{$^{1}$Institute of Physics, Chinese Academy of Sciences, P.O. Box 603, 
Beijing 100190, China\\
$^{2}$Texas Center for Superconductivity, University of Houston, Houston, Texas 77204, USA}
 
\date{\today}

\begin{abstract}
By taking into account the possibility of all the intralayer as well as the interlayer current orderings, we derive an eight-band model for interacting electrons in bilayer graphene. With the numerical solution to the model, we show that only the current orderings between the same sublattice sites can exist within the range of the physical interacting strength. This result confirms our previous model of spin-polarized-current phase for the ground-state of interacting electrons in bilayer graphene that resolves a number of experimental puzzles.   
\end{abstract}


\maketitle

\section{Introduction}

Bilayer graphene (BLG) has been attracted much attention because of its potential application to new electronic devices.\cite{Ohta,Oostinga,McCann,Castro} The experimental observations on high quality suspended BLG samples \cite{Weitz,Freitag,Velasco,Bao,Elferen,Velasco1} show that the ground state of the electrons at the charge neutrality point (CNP) is insulating with puzzling properties in the presence of external magnetic field: (i) the insulating gap can be closed by a perpendicular electric field of either polarity,\cite{Velasco} (ii) the gap grows greatly with increasing magnetic field $B$ about 46 times larger than the Zeeman splitting at $B > 0.1$ Tesla,\cite{Velasco} (iii) the state is particle-hole asymmetric in the presence of the magnetic field,\cite{Velasco} (iv) there is a peak structure in the electric conductivity at small $B$ and at the CNP,\cite{Weitz} and (v) there are quantum Hall states at all integers from 0 to $\pm 4$.\cite{Elferen,Velasco1} On theoretical aspect, a number of models for the ground state of the electrons in BLG have been proposed, which include the ferroelectric-layer asymmetric state \cite{Min,Nandkishore} or quantum valley Hall state,\cite{Zhang2} a layer-polarized antiferromagnetic state,\cite{Nilsson,Gorbar} a quantum anomalous Hall state,\cite{Jung,Nandkishore1,Zhang1} a quantum spin Hall state,\cite{Jung,Zhang1} a superconducting state in coexistence with antiferromagnetism,\cite{Milovanovic} an ordered-current state,\cite{Zhu,Yan1,Yan2}, and a spin-polarized current state (SPCS).\cite{Yan3} Besides these, the insulating gap has also been explained as electron scattering on boundaries between domains with different stacking (AB and BA).\cite{san-jose} The earlier studies based on the renormalization group approach predicted the Lifshitz transition in the interacting electron system of BLG.\cite{Lemonik} Amongst these models, only the SPCS can qualitatively explain the above experimental observations. 

By the SPCS model,\cite{Yan3} there are spin-dependent currents between the lattice sites due to the interactions between electrons. In our previous work, only the intra-sublattice current orders were taken into account. One then raises a question: are there any possible inter-sublattice current orderings? In this work, we will answer this question by deriving and solving an eight-band model that counts in all the possible current orderings of interacting electrons in the BLG.

\section{Formalism}

The Hamiltonian of the electrons is
\begin{eqnarray}
H&=&-\sum_{ij\sigma}t_{ij}c^{\dagger}_{i\sigma}c_{j\sigma} +\frac{1}{2}\sum_{ij}v_{ij}\delta n_{i}\delta n_{j} \label {hm}
\end{eqnarray}
where $c^{\dagger}_{i\sigma} (c_{i\sigma})$ creates (annihilates) an electron of spin $\sigma$ at site $i$, $t_{ij}$ is the hopping energy between sites $i$ and $j$, $\delta n_{i}=n_i-n$ is the number deviation of electrons at site $i$ from the average occupation $n$, and $v$'s are the interactions between electrons. We use the tight-binding model considering only the intra-layer NN [between $a$ ($a'$) and $b$ ($b'$)] and interlayer NN (between $b$ and $a'$) electron hoppings given by $t$ = 3 eV and $t_1$ = 0.273 eV, respectively.\cite{Tatar,LMZ} We will use the mean-field approximation (MFA) and therefore adopt effective interactions that include screening due to the electronic charge fluctuations. As in our previous model,\cite{Yan3} the effective interactions are given as
\begin{eqnarray}
v_{ij} = v(r) = \frac{e^2}{r(1+ 4.69 q_sr+q^2_s r^2)} \label{int}
\end{eqnarray}  
where $r = |\vec r|$ with $\vec r$ as a vector from site $i$ to site $j$, and $q_s = 2\pi e^2\chi_0$ is the screening constant with $\chi_0 = t_1\ln 4/\pi(a\epsilon_0)^2$ (and $\epsilon_0 = \sqrt{3}t/2$) the polarizability by the random-phase-approximation (RPA).\cite{Hwang} The form of $v(r)$ is consistent with the RPA in the limit $r \to \infty$. In our previous work,\cite{Yan3} we have shown that except the intra-sublattice current ordering there are no charge and spin orderings with absence of the magnetic field. We here confine ourselves to the case of no external magnetic field and at the CNP. We therefore ignore the on-site interaction and simplify the discussion only on the current orderings.

First, we explain why the current orderings may appear in the system. By the MFA, the interaction term in Eq. (\ref{hm}) is approximated as
\begin{eqnarray}
\frac{1}{2}\sum_{ij}v_{ij}\delta n_{i}\delta n_{j} \approx \sum_{ij\sigma}v_{ij}\langle c_{i\sigma}c^{\dagger}_{j\sigma}\rangle c^{\dagger}_{i\sigma}c_{j\sigma}. \label {inthm}
\end{eqnarray}
In general, the average $\langle c_{i\sigma}c^{\dagger}_{j\sigma}\rangle$ is not always real, but may contain an imaginary part,\cite{Varma}
\begin{eqnarray}
\langle c_{i\sigma}c^{\dagger}_{j\sigma}\rangle = R_{ij\sigma}+iI_{ij\sigma}. \label {av}
\end{eqnarray}
The imaginary part $I_{ij\sigma}$ here corresponds to a current, which is self-consistently determined by the MFA. The treatment of the current between the sites of same sublattice has been detailed in Ref. \onlinecite{Yan2}.

In addition to the intra-sublattice currents, we here consider the inter-sublattice currents between the sites in the same layer as well as in different layers. First, we analyze the currents between the two sublattices $a$ and $b$, for example, on the top layer shown in Fig. 1. The consideration applies to the currents between other sublattices. In Fig. 1, we draw all the nearest-neighbor (NN) bond currents between the $a$ (black) and $b$ (white) atoms. The purple lines denote the current from atom $b$ to atom $a$, the red and green lines are respectively the left-turn and right-turn currents from $a$ to $b$. The hexagons consisting of green and purple bonds contain the right-turn current loops, while the hexagons of red and purple bonds bear the left-turn current loops. The current in the purple bond equals to the sum of the currents in the red and green bonds because of the current continuity. In this process, the current density at each atom does not vanish, which is different from the case of the currents between the same sublattice where the current density vanishes. Since currents are in loop forms, the total current is zero. 

\begin{figure}
\centerline{\epsfig{file=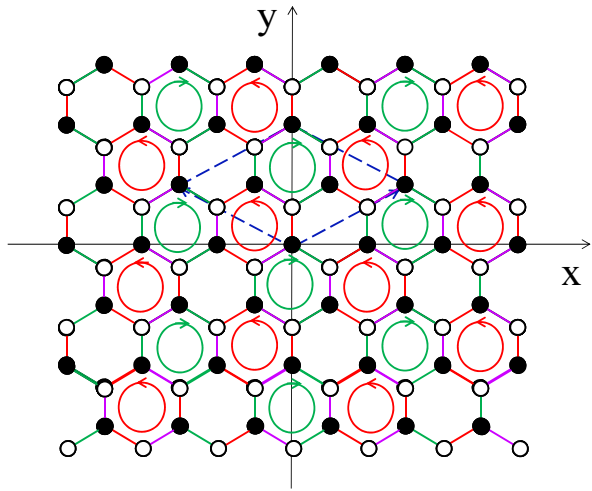,height=7 cm}}
\caption{(Color online) Bond currents between atoms $a$ (black) and $b$ (white) in top layer of BLG. The purple lines denote the current from atom $b$ to atom $a$, the red and green lines are respectively the left-turn and right-turn currents from $a$ to $b$. The red and green circles with arrow indicate the circulations of the current in the hexagon bonds. The diamond area enclosed by the dashed lines is the unit cell.} 
\end{figure} 

From Fig. 1, we see that the translational invariance of the system is broken. One then needs a superlattice with unit cell containing more $a$ and $b$ atoms to describe the electrons. We find that there are three $a$ atoms and three $b$ atoms in the unit cell as enclosed by the dashed lines in Fig. 1 for a single layer. In Fig. 2, we depict the unit cell for the BLG consisting of three $a$ atoms $a_1$, $a_2$ and $a_3$ and three $b$ atoms $b_1$, $b_2$ and $b_3$ on the top layer and three $a'$ atoms $a'_1$, $a'_2$ and $a'_3$ and three $b'$ atoms $b'_1$, $b'_2$ and $b'_3$ on the bottom layer. From the top view, the $b$ atoms are just on top of the $a'$ atoms, while the $a$ atoms are at the hexagon center of the bottom atoms. The lattice constant defined as the distance between the NN same atoms (for example $a$ atoms) is $a \approx 2.4$ \AA~ and interlayer distance $d \approx 3.34$ \AA. We will use the unit $a$ = 1 for the length. We take the basis of the unit cell as 
\begin{eqnarray}
\vec e_1 &=& (3/2,\sqrt{3}/2) \nonumber\\
\vec e_2 &=& (-3/2,\sqrt{3}/2) \nonumber
\end{eqnarray}
where the numbers in the parenthesis are the $x$ and $y$ coordinates. The area of the unit cell is $S = |\vec e_1 \times \vec e_2| = 3\sqrt{3}/2$. The basis of Brillouin zone (BZ) in the reciprocal space are given as
\begin{eqnarray}
\vec v_1 &=& \vec e_2\times \hat z/S = (1/3,1/\sqrt{3}) \label{v1}\\
\vec v_2 &=& \hat z\times\vec e_1 /S = (-1/3,1/\sqrt{3}) \label{v2}
\end{eqnarray}
where $\hat z$ is a unit vector perpendicularly pointing out of the plane. The reciprocal lattice points $2n_1\pi \vec v_1 + 2n_2\pi\vec v_2$ with $n_1$ and $n_2$ integers and the first BZ are shown in Fig. 3. Note that the Dirac points at the corners of the original first BZ as shown in Fig. 3 are now the lattice points of the reciprocal space. Therefore, the zero energy of noninteracting electrons now appears at the origin since the energy is a periodic function of momentum in the reciprocal space of the superlattice. 

\begin{figure} 
\centerline{\epsfig{file=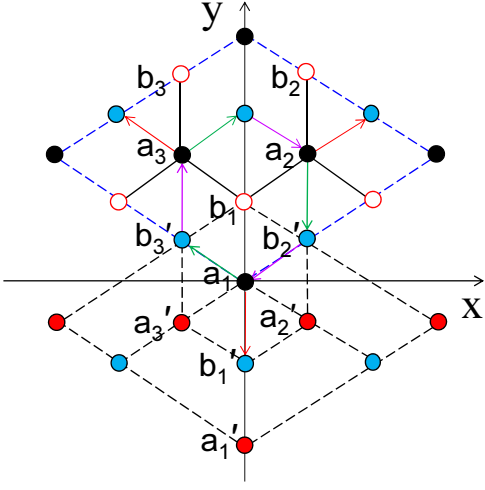,width=7. cm}}
\caption{(Color online) Unit cell of the bilayer graphene containing three $a$ atoms (black) $a_1$, $a_2$ and $a_3$ and three $b$ atoms (white with red edge) $b_1$, $b_2$ and $b_3$ on the top layer and the corresponding $a'$ (red) and $b'$ (blue) atoms on the bottom layer. The arrows denote the interlayer currents between the $a$ and $b'$ atoms.} 
\end{figure} 

We will still use the term, sublattice, to distinguish the atoms in the original BLG lattice. So, there are four sublattices $a$, $b$, $a'$, and $b'$. However, for the superlattice, there are 12 atoms in each unit cell. The term `subspace' will be used for describing a space consisting of one kind of these 12 atoms; there are 12 subspaces in the system.  

\subsection{12-band model}

For description of the problem in momentum space, we first define the operator in real space
\begin{eqnarray}
\psi^{\dagger}_{j\sigma} = (\psi^{\dagger}_{j\sigma 1},\psi^{\dagger}_{j\sigma 2}) \label{op}
\end{eqnarray}  
with 
\begin{eqnarray}
\psi^{\dagger}_{j\sigma 1} &=& (a^{\dagger}_{j1\sigma},b^{\dagger}_{j1\sigma},a^{\dagger}_{j2\sigma},b^{\dagger}_{j2\sigma},a^{\dagger}_{j3\sigma},b^{\dagger}_{j3\sigma})\label{op1} \\
\psi^{\dagger}_{j\sigma 2} &=& (a'^{\dagger}_{j1\sigma},b'^{\dagger}_{j1\sigma},a'^{\dagger}_{j2\sigma},b'^{\dagger}_{j2\sigma},a'^{\dagger}_{j3\sigma},b'^{\dagger}_{j3\sigma})\label{op2}  
\end{eqnarray}  
where $a^{\dagger}_{j\mu\sigma}$ ($b^{\dagger}_{j\mu\sigma}$) creates an electron of spin $\sigma$ at $\mu$th atom $a$ ($b$) in $j$th unit cell. Then, the operator in momentum space is given by
\begin{eqnarray}
\psi_{k\sigma} = \frac{1}{\sqrt{N}}\sum_j\psi_{j\sigma}\exp(-i\vec k\cdot \vec j) \label{ft}
\end{eqnarray}  
where $N$ is total number of the unit cells in the superlattice, and the momentum $k$ is confined within the first BZ enclosed by the dashed line in Fig. 3. For the sake of description, we name an $\mu$th atom of sublattice $l$ in unit cell of the superlattice as the $l_{\mu}$ atom. For example, $a^{\dagger}_{k2\sigma}$ creates an electron of momentum $k$ and spin $\sigma$ at $a_2$ atom.

\begin{figure}[b] 
\centerline{\epsfig{file=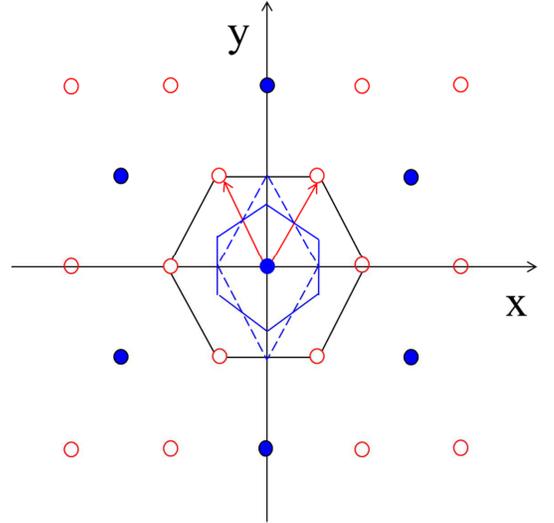,width=7.5cm}}
\caption{(Color online) Lattice points in the reciprocal space. The solid blue points are for the original honeycomb lattice with the area enclosed by the black solid line as the first Brillouin zone (BZ). The red-edge circles are the additional points for the superlattice with unit cell containing three $a$-atoms and three $b$-atoms. The two vectors represent the basis of the reciprocal lattice. The diamond area with dashed-line edge is the first BZ for the superlattice. The inner hexagon is the equivalent high symmetry first BZ.} 
\end{figure} 

In the momentum-space representation, we obtain the expression for the non-interacting Hamiltonian as
\begin{eqnarray}
H_0 =  \sum_{k\sigma}\psi^{\dagger}_{k\sigma}H(k)\psi_{k\sigma} \label {hm0}
\end{eqnarray}
with
\begin{eqnarray}
H(k)=\begin{pmatrix}
H_s&C\\
C^{\dagger}&H_s \\
\end{pmatrix}\label{hk}\\
C=-t_1\begin{pmatrix}
\sigma_-&0&0\\
0&\sigma_-&0\\
0&0&\sigma_-\\
\end{pmatrix}\nonumber
\end{eqnarray}
and
\begin{eqnarray}
H_s=-t\begin{bmatrix}
0& 1&0&e^{-i(k_1+k_2)}&0&e^{-i(k_1+k_2)}\\
1&0&1&0&1&0 \\
0& 1&0&1&0&e^{-ik_2}\\
e^{i(k_1+k_2)}&0& 1&0&e^{ik_1}&0\\
0&1&0&e^{-ik_1}&0&1\\
e^{i(k_1+k_2)}&0&e^{ik_2}&0&1&0\\
\end{bmatrix}\nonumber
\end{eqnarray}
where $\sigma_-$ is the Pauli matrix, $k_1$ and $k_2$ are the momentum components along $\vec v_1$ and $\vec v_2$, respectively. By the MFA for the interactions, the element of the self-energy is obtained as
\begin{eqnarray}
\Sigma^{ll',\sigma}_{\mu\nu}(k) = -\frac{1}{N}\sum_{k'}v^{ll'}_{\mu\nu}(|\vec k-\vec k'|)(\langle c^{\dagger}_{k'l'\nu\sigma}c_{k'l\mu\sigma}\rangle -\delta_{l_{\mu}l'_{\nu}}/2)  \nonumber
\end{eqnarray}
where $c^{\dagger}_{kl\mu\sigma}$ ($c_{kl\mu\sigma}$) creates (annihilates) an electron of momentum $k$ and spin $\sigma$ at $l_{\mu}$ subspace, and $v^{ll'}_{\mu\nu}(q)$ is the lattice Fourier component of the interaction,
\begin{eqnarray}
v^{ll'}_{\mu\nu}(q) = \sum_{\vec R}v(|\vec R + \vec r^{ll'}_{\mu\nu}|)\exp(-i\vec q\cdot\vec R)  \label {vq}
\end{eqnarray}
with $\vec r^{ll'}_{\mu\nu}$ the vector from the $l'_{\nu}$ atom to the $l_{\mu}$ atom in the unit cell and the $\vec R$-summation running over the positions of the unit cells in the superlattice. The element of the effective Hamiltonian is $H^{ll'}_{\mu\nu}(k)+ \Sigma^{ll',\sigma}_{\mu\nu}(k)$. Since there are 12 atoms in the unit cell of the superlattice, we thus have a 12-band model for the electrons. 

For the carrier concentration close to the CNP, only the energy levels close to zero momentum need to be considered. We here consider the system at the CNP. In our previous work, the elements of the self-energy at $k = 0$ have been proved to be pure imaginary numbers that come from the current orderings for the system at the CNP. We therefore have
\begin{eqnarray}
\Sigma^{ll',\sigma}_{\mu\nu}(0) &=& -\frac{i}{N}\sum_{k'}v^{ll'}_{\mu\nu}(k'){\rm Im}\langle c^{\dagger}_{k'l'\nu\sigma}c_{k'l\mu\sigma}\rangle  \nonumber\\
&\approx& -i\frac{v^{ll'}_{\mu\nu}(0)}{N}\sum_{k'}{\rm Im}\langle c^{\dagger}_{k'l'\nu\sigma}c_{k'l\mu\sigma}\rangle  \label{se1}
\end{eqnarray}
where the second line comes from the fact that the average ${\rm Im}\langle c^{\dagger}_{k'l'\nu\sigma}c_{k'l\mu\sigma}\rangle$ is sizable only when $k'$ is close to 0. To understand the physics of such an element of the self-energy, we consider the current 
\begin{eqnarray}
I^{ll',\sigma}_{\mu\nu} &=& {\rm Im} \langle c^{\dagger}_{j'l'\nu\sigma}c_{jl\mu\sigma}\rangle   \nonumber\\
&=& \frac{1}{N}\sum_{k}{\rm Im}[\langle c^{\dagger}_{kl'\nu\sigma}c_{kl\mu\sigma}\rangle e^{i\vec k\cdot(\vec j-\vec j')}]
 \nonumber\\
&\approx& \frac{1}{N}\sum_{k}{\rm Im}\langle c^{\dagger}_{kl'\nu\sigma}c_{kl\mu\sigma}\rangle    \nonumber
\end{eqnarray}
where the same approximation as in Eq. (\ref{se1}) again has been used in the last line for not too large $|\vec j-\vec j'|$. But the equality in the last line is valid for the $l_{\mu}$- and $l'_{\nu}$ atoms belonging to the same unit cell. Therefore, the self-energy can be expressed as
\begin{eqnarray}
\Sigma^{ll',\sigma}_{\mu\nu}(0) \equiv \Sigma^{ll',\sigma}_{\mu\nu} \approx -i v^{ll'}_{\mu\nu}(0)I^{ll',\sigma}_{\mu\nu}.  \label{se2}
\end{eqnarray}
Note that $I^{ll,\sigma}_{\mu\mu} = 0$. The remaining 6$\times$11 current elements are not independent quantities, but satisfy the current continuity law and the BLG lattice symmetry. We analyze the current orders below.

(1) Currents between the atoms in the same sublattice. For this case, we have obtained the result in our previous works,\cite{Yan2,Yan3} 
\begin{eqnarray}
I^{ll,\sigma}_{\mu\nu} = I^{l,\sigma}\sin(\vec K\cdot \vec r^{ll}_{\mu\nu})  \label{iaa}
\end{eqnarray}
where $\vec K = (4\pi/3,0)$ is a Dirac point, and $I^{l,\sigma}$ is a constant satisfying $I^{a,\sigma} = -I^{b',\sigma}$ and $I^{b,\sigma} = -I^{a',\sigma}$. Since $\vec K$ is a lattice point in the reciprocal space, the current $I^{ll,\sigma}_{\mu\nu}$ is a periodic function in the superlattice. It means that the current $I^{ll,\sigma}_{\mu\nu}$ is the same when changing the position of the $\mu$th atom to another unit cell but with $\nu$th atom fixed. Therefore, the formula given by Eq. (\ref{iaa}) represents not only the currents between the same type atoms within the unit cell but also between that in different cells. Figure 4 shows the currents flow through an atom from/to the same type atoms on a layer. The current density in this case is zero at every atom. All the bond currents between the same type atoms constitute current loops in the lattice. Instead of considering $I^{l,\sigma}$, we define the order parameter 
\begin{eqnarray}
\Delta^{l}_{\sigma} = -\sqrt{3}v^{ll}_{12}(0)I^{ll,\sigma}_{12}.  \label{daa}
\end{eqnarray}
We then have
\begin{eqnarray}
\Sigma^{ll,\sigma}_{12}
= -\Sigma^{ll,\sigma}_{13} 
= \Sigma^{ll,\sigma}_{23} = i \Delta^{l}_{\sigma}/\sqrt{3}. 
\label{saa}
\end{eqnarray}
The relationship given by Eq. (\ref{saa}) can be obtained from Eqs. (\ref{se2}) and (\ref{iaa}).

\begin{figure}
\centerline{\epsfig{file=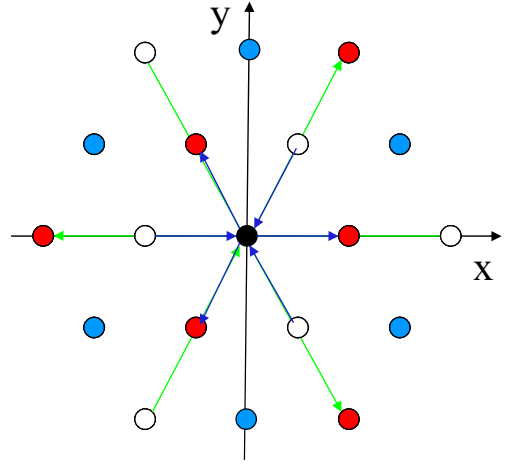,height=7 cm}}
\caption{(Color online) Currents through the central atom (black) from/to the same type atoms on a layer. The currents are incoming (outgoing) from (to) the white (red) atoms. There are no currents between the black and the blue sites.} 
\end{figure} 

(2) Currents between atoms in different sublattices. By viewing Fig. 1, we can obtain the formula for the bond currents between the sublattices $a$ and $b$. The result can be extended to the case of currents between two atoms of any other two sublattices $l$ and $l'$ except $ll'= ab'$ and $ll' = ba'$. For $ll' \ne ab'$ and $ba'$, we have 
\begin{eqnarray}
I^{ll',\sigma}_{\mu\nu} = I^{ll',\sigma}\cos(\vec K_{\nu}\cdot \vec r^{ll'}_{\mu\nu}),  \label{iab}
\end{eqnarray}
where $I^{ll',\sigma}$ is a constant, and $\vec K_{\nu}$'s are three Dirac points defined as
\begin{eqnarray}
\vec K_1 &=&(-2\pi/3,2\pi/\sqrt{3}) \nonumber\\
\vec K_2 &=& (4\pi/3,0)  \nonumber\\
\vec K_3 &=& (2\pi/3,2\pi/\sqrt{3}).  \nonumber
\end{eqnarray}
One can check that the currents given by Eq. (\ref{iab}) satisfy the continuity law. Again, the $I^{ll',\sigma}_{\mu\nu}$ is a periodic function in the superlattice. In Fig. 5, some other bond currents between the atoms in different unit cells are also depicted, which give rise to the larger current loops. For $ll' = ab'$, the bond currents between $a$ and $b'$ atoms are depicted in Fig. 2. We have 
\begin{eqnarray}
I^{ab',\sigma}_{\mu\nu} = I^{ab',\sigma}\cos(\vec Q_{\nu}\cdot \vec r^{ab'}_{\mu\nu}),  \label{iabp}
\end{eqnarray}
where $\vec Q_1 = \vec K_1$, $\vec Q_2 = \vec K_3$, and $\vec Q_3 = \vec K_2$. 
For the currents between the sublattices $b$ and $a'$, since the two sublattices look like a same sublattice from top view, we obtain the formula as
\begin{eqnarray}
I^{ba',\sigma}_{\mu\nu} = I^{ba',\sigma}\sin(\vec K\cdot \vec r^{ba'}_{\mu\nu}).  \label{ibap}
\end{eqnarray}
Define the order parameters
\begin{eqnarray}
\Delta^{ll'}_{\sigma} &=& -3v^{ll'}_{11}(0)I^{ll',\sigma}_{11}, ~~ll' \ne ba' \label{doff} \\
\Delta^{ba'}_{\sigma} &=& -\sqrt{3}v^{ba'}_{12}(0)I^{ba',\sigma}_{12}.  \label{dba}
\end{eqnarray}
From Eqs. (\ref{se2}), (\ref{iab}), (\ref{doff}), and (\ref{dba}), we can determine the relationship between the elements of the self-energy. For example, we have
\begin{eqnarray}
 \Delta^{ab}_{\sigma}/3 &=& \Sigma^{ab,\sigma}_{11}
= \Sigma^{ab,\sigma}_{12}= -\Sigma^{ab,\sigma}_{13}/2 \nonumber\\
&=& \Sigma^{ab,\sigma}_{21}= -\Sigma^{ab,\sigma}_{22}/2 = \Sigma^{ab,\sigma}_{23} \nonumber\\
&=& -\Sigma^{ab,\sigma}_{31}/2 = \Sigma^{ab,\sigma}_{32} = \Sigma^{ab,\sigma}_{33} 
\label{sab}
\end{eqnarray}
and $\Sigma^{ba,\sigma}_{\nu\mu}=-\Sigma^{ab,\sigma}_{\mu\nu}$. Here, we have used the fact that $v^{ab}_{\mu\nu}(0) = v^{ab}_{11}(0)$ since $\vec r^{ab}_{\mu\nu}$ appearing in the definition of $v^{ab}_{\mu\nu}(0)$ given by Eq. (\ref{vq}) is a shift of a superlattice constant from $\vec r^{ab}_{11}$ or in addition then a rotation of angle $\pm 2\pi/3$; because the lattice is invariant under rotations $\pm 2\pi/3$, these operations on $\vec r^{ab}_{11}$ do not change the result of the $\vec r$-summation. The consideration applies to other interactions. We therefore consider only the interactions $v^{ll'}_{\mu\nu}(0)$ with indices for the atoms $a_1$, $b_1$, $b'_1$ and the $a'$ atom just beneath $b_1$ as show in Fig. 2.

\begin{figure}
\centerline{\epsfig{file=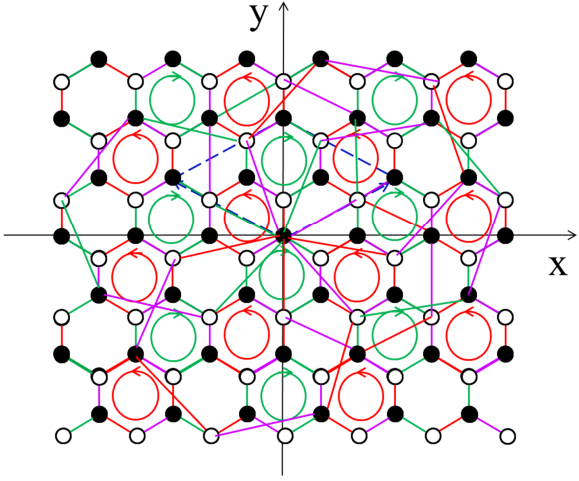,height=7 cm}}
\caption{(Color online) The same as Fig. 1 but with some currents of longer bonds connected to the original atom depicted.} 
\end{figure} 

All the above results can be summarized in the self-energy matrix
\begin{eqnarray}
\Sigma^{\sigma} = \begin{pmatrix}
\Sigma_t&\Sigma_{tb}\\
\Sigma^{\dagger}_{tb}&\Sigma_b \\
\end{pmatrix}           \label{tsf}
\end{eqnarray}
where $\Sigma_t$ and $\Sigma_b$ are self-energy matrices for the top and bottom layers, respectively, and $\Sigma_{tb}$ is the self-energy matrix due to the interlayer currents. $\Sigma_t$ is given by 
\begin{equation}
\Sigma_t=\frac{i}{3}\begin{pmatrix}
0&\Delta_{\sigma}^{ab}&\tilde\Delta_{\sigma}^{a}&\Delta_{\sigma}^{ab}&-\tilde\Delta_{\sigma}^{a}&-2\Delta_{\sigma}^{ab}\\
-\Delta_{\sigma}^{ab}&0&-\Delta_{\sigma}^{ab}&\tilde\Delta_{\sigma}^{b}&2\Delta_{\sigma}^{ab}&-\tilde\Delta_{\sigma}^{b}\\
-\tilde\Delta_{\sigma}^{a}&\Delta_{\sigma}^{ab}&0&-2\Delta_{\sigma}^{ab}&\tilde\Delta_{\sigma}^{a}&\Delta_{\sigma}^{ab}\\
-\Delta_{\sigma}^{ab}&-\tilde\Delta_{\sigma}^{b}&2\Delta_{\sigma}^{ab}&0&-\Delta_{\sigma}^{ab}&\tilde\Delta_{\sigma}^{b}\\
\tilde\Delta_{\sigma}^{a}&-2\Delta_{\sigma}^{ab}&-\tilde\Delta_{\sigma}^{a}&\Delta_{\sigma}^{ab}&0&\Delta_{\sigma}^{ab}\\
2\Delta_{\sigma}^{ab}&\tilde\Delta_{\sigma}^{b}&-\Delta_{\sigma}^{ab}&-\tilde\Delta_{\sigma}^{b}&-\Delta_{\sigma}^{ab}&0\\
\end{pmatrix}\nonumber
\end{equation}
with $\tilde\Delta^a_{\sigma} = \sqrt{3}\Delta^a_{\sigma}$ and $\tilde\Delta^b_{\sigma} = \sqrt{3}\Delta^b_{\sigma}$. The expression for $\Sigma_b$ is obtained by the replacements $\tilde\Delta_{\sigma}^a \to -\tilde\Delta_{\sigma}^b$, $\tilde\Delta_{\sigma}^b \to -\tilde\Delta_{\sigma}^a$, and $\Delta_{\sigma}^{ab} \to \Delta_{\sigma}^{a'b'}$ in the above matrix $\Sigma_t$. For $\Sigma_{tb}$, we have
\begin{equation}
\setlength{\arraycolsep}{0.1pt}
\Sigma_{tb}=\frac{i}{3}\begin{bmatrix}
\Delta_{\sigma}^{aa'}&\Delta_{\sigma}^{ab'}&\Delta_{\sigma}^{aa'}&-2\Delta_{\sigma}^{ab'}&-2\Delta_{\sigma}^{aa'}&\Delta_{\sigma}^{ab'}\\
0&\Delta_{\sigma}^{bb'}&\tilde\Delta_{\sigma}^{ba'}&\Delta_{\sigma}^{bb'}&-\tilde\Delta_{\sigma}^{ba'}&-2\Delta_{\sigma}^{bb'}\\
\Delta_{\sigma}^{aa'}&-2\Delta_{\sigma}^{ab'}&-2\Delta_{\sigma}^{aa'}&\Delta_{\sigma}^{ab'}&\Delta_{\sigma}^{aa'}&\Delta_{\sigma}^{ab'}\\
-\tilde\Delta_{\sigma}^{ba'}&\Delta_{\sigma}^{bb'}&0&-2\Delta_{\sigma}^{bb'}&\tilde\Delta_{\sigma}^{ba'}&\Delta_{\sigma}^{bb'}\\
-2\Delta_{\sigma}^{aa'}&\Delta_{\sigma}^{ab'}&\Delta_{\sigma}^{aa'}&\Delta_{\sigma}^{ab'}&\Delta_{\sigma}^{aa'}&-2\Delta_{\sigma}^{ab'}\\
\tilde\Delta_{\sigma}^{ba'}&-2\Delta_{\sigma}^{bb'}&-\tilde\Delta_{\sigma}^{ba'}&\Delta_{\sigma}^{bb'}&0&\Delta_{\sigma}^{bb'}\\
\end{bmatrix},\nonumber
\end{equation}
where $\tilde\Delta_{\sigma}^{ba'}= \sqrt{3}\Delta_{\sigma}^{ba'}$.
 
\subsection{8-band model}

To get an effective low energy Hamiltonian, we first expand the noninteracting Hamiltonian matrix $H_s$ at $k = 0$ to linear $k$,
\begin{eqnarray}
H_s=h^0 + h^1_k \nonumber
\end{eqnarray}
with 
\begin{eqnarray}
h^0&=&-t\begin{pmatrix}
\sigma_1& \sigma_1&\sigma_1\\
\sigma_1& \sigma_1&\sigma_1\\
\sigma_1& \sigma_1&\sigma_1\\
\end{pmatrix},\nonumber\\
\setlength{\arraycolsep}{0.2pt}
h^1_k&=&-it\begin{bmatrix}
0& -(k_1+k_2)\sigma_+&-(k_1+k_2)\sigma_+\\
(k_1+k_2)\sigma_-& 0&k_1\sigma_--k_2\sigma_+\\
(k_1+k_2)\sigma_-&-k_1\sigma_++k_2\sigma_-&0\\
\end{bmatrix}.\nonumber
\end{eqnarray}
The matrix $h^0$ can be written as $h^0 = -tM\otimes\sigma_1$ with $\otimes$ as the Kronecker product and $M$ is a 3$\times$3 matrix with all the elements = 1. The eigenvalues of $M$ are 0 with dually degeneracy and 3. From the corresponding eigen-functions of these eigenvalues, we define the matrix
\begin{eqnarray}
T=\frac{1}{\sqrt{3}}\begin{pmatrix}
e^{i\alpha}& 1&1\\
e^{-i\alpha}&e^{-i\alpha}&1\\
1& e^{i\alpha}&1\\
\end{pmatrix}\otimes\sigma_0\nonumber
\end{eqnarray}
with $\alpha = 2\pi/3$. With this matrix, we do transformation of $H_s$
\begin{eqnarray}
T^{\dagger}H_sT=\epsilon_0\begin{pmatrix}
H_0 & F\\
F^{\dagger}&D\\
\end{pmatrix}\label{tr1}
\end{eqnarray}
with $\epsilon_0 = \sqrt{3}t/2$, and 
\begin{eqnarray}
H_0&=&\begin{pmatrix}
p_x\sigma_1+p_y\sigma_2& 0\\
0&-p_x\sigma_1+p_y\sigma_2\\
\end{pmatrix},\label{h0}\\
F&=&\begin{pmatrix}
-3(p_x-ip_y)e^{-i\alpha}\sigma_1\\
3(p_x+ip_y)\sigma_1\\
\end{pmatrix},\label{fm}\\
D&=&-2\sqrt{3}\sigma_1-p_y\sigma_2,\label{dm}
\end{eqnarray}
where $p_x$ and $p_y$ are the momentum components in the $x-y$ orthogonal system
\begin{eqnarray}
p_x &=& (k_1-k_2)/3 \nonumber\\
p_y &=& (k_1+k_2)/\sqrt{3}. \nonumber
\end{eqnarray}
Clearly, $H_0$ describes the energy bands close to zero with the two diagonal matrices expressing the low-energy non-interacting electrons in the two valleys, while $D$ gives rise to the bands with energies about $\pm 2\sqrt{3}\epsilon_0$ far from the zero. The low energy band of $D$ is fully occupied while the high energy band is empty. There is no contribution to the physical process from these two bands. These two bands will be ignored. The off-diagonal matrices $F$ and $F^{\dagger}$ are high order corrections. To see it, we take further transformation of the right-hand side of Eq. (\ref{tr1}) using
\begin{eqnarray}
T_1 = \begin{pmatrix}
I_4 & 0\\
-D^{-1}F^{\dagger}&\sigma_0\\
\end{pmatrix}\nonumber
\end{eqnarray}
with $I_4$ as the 4$\times$4 unit matrix and obtain
\begin{eqnarray}
T^{\dagger}_1\begin{pmatrix}
H_0 & F\\
F^{\dagger}&D\\
\end{pmatrix}T_1 = \begin{pmatrix}
H_0-FD^{-1}F^{\dagger} & 0\\
0&D\\
\end{pmatrix}.\nonumber
\end{eqnarray}
The correction $FD^{-1}F^{\dagger}$ is a second order in the momentum, and will be neglected. 

The Hamiltonian $H(k)$ given by Eq. (\ref{hk}) and the self-energy $\Sigma^{\alpha}$ given by Eq. (\ref{tsf}) are transformed by $I_2\otimes T$ with $I_2$ as the 2$\times$2 unit matrix operating in the space of top and bottom layers. The matrix $C$ in $H(k)$ is unchanged under the transformation. The self-energy matrices are transformed as
\begin{eqnarray}
T^{\dagger}\Sigma_tT&=&\begin{pmatrix}
-m_t&\Delta_{\sigma}^{ab}\sigma_2 & 0\\
\Delta_{\sigma}^{ab}\sigma_2&m_t&0\\
0&0&0&\\
\end{pmatrix},\label{tst}\\
T^{\dagger}\Sigma_bT&=&\begin{pmatrix}
-m_b&\Delta_{\sigma}^{a'b'}\sigma_2 & 0\\
\Delta_{\sigma}^{a'b'}\sigma_2&m_b&0\\
0&0&0&\\
\end{pmatrix},\label{bst}\\
T^{\dagger}\Sigma_{tb}T&=&\begin{pmatrix}
-\Delta_{\sigma}^{ba'}\sigma_-&-im_{\alpha} & 0\\
-im_{-\alpha}&\Delta_{\sigma}^{ba'}\sigma_-&0\\
0&0&0&\\
\end{pmatrix},\label{tbst}
\end{eqnarray}
with 
\begin{eqnarray}
m_t&=&\begin{pmatrix}
\Delta_{\sigma}^{a}&0\\
0&\Delta_{\sigma}^{b}\\
\end{pmatrix}, ~~~~
m_b=-\begin{pmatrix}
\Delta_{\sigma}^{b}&0\\
0&\Delta_{\sigma}^{a}\\
\end{pmatrix},\nonumber\\
m_{\alpha}&=&\begin{pmatrix}
\Delta_{\sigma}^{aa'}&\Delta_{\sigma}^{ab'}e^{i\alpha}\\
0&\Delta_{\sigma}^{bb'}\\
\end{pmatrix}. \nonumber
\end{eqnarray}
The zeros in the third row and third column of matrices in Eqs. (\ref{tst})-(\ref{tbst}) stem from the current continuity law that results in zero of the total current.

\begin{widetext}
By keeping only the energy bands close to zero, we get an effective eight-band Hamiltonian as
\begin{eqnarray}
H^{eff}(p)=&\begin{pmatrix}
p_x\sigma_1+p_y\sigma_2- m_t&\Delta_{\sigma}^{ab}\sigma_2& -(t_1+\Delta_{\sigma}^{ba'} )\sigma_-&-i m_{\alpha}\\
\Delta_{\sigma}^{ab}\sigma_2&-p_x\sigma_1+p_y\sigma_2+ m_t& -i m_{-\alpha}&-(t_1-\Delta_{\sigma}^{ba'})\sigma_-\\
-(t_1+\Delta_{\sigma}^{ba'})\sigma_+& i m^{\dagger}_{-\alpha}&p_x\sigma_1+p_y\sigma_2-m_b&\Delta_{\sigma}^{a'b'}\sigma_2\\
i m^{\dagger}_{\alpha}&-(t_1-\Delta_{\sigma}^{ba'})\sigma_+& \Delta_{\sigma}^{a'b'}\sigma_2
&-p_x\sigma_1+p_y\sigma_2+ m_b\\
\end{pmatrix},
\label{heff}
\end{eqnarray}
\end{widetext}
where we have set the unit of energy as $\epsilon_0$ = 1. Note that by setting all inter-sublattice current orders as zero, the formula of $H^{eff}$ can be transformed to a matrix with two $4\times$4 matrices in the diagonal. The two $4\times$4 matrices represent respectively the four-band Hamiltonian in the two valleys. Therefore, equation (\ref{heff}) reduces to our previous result. From Eq. (\ref{heff}), it is clear that the current orders $\Delta_{\sigma}^{ab}$ and $\Delta_{\sigma}^{a'b'}$ give rise to the inter-valley couplings. This is because that the momentum of an electron in these current processes changes as its moving direction changes. We can neglect current order $\Delta_{\sigma}^{ba'}$ since it is a correction much smaller than the hopping $t_1$ between $b$ and $a'$. 

With the effective Hamiltonian, we then find out the eigen-energy $E_{\lambda}(p)$ and the wavefunction $\phi_{\mu\lambda}(p)$ by solving the equation 
\begin{eqnarray}
\sum_{\nu}H^{eff}_{\mu\nu}(p)\phi_{\nu\lambda}(p)=E_{\lambda}(p)\phi_{\mu\lambda}(p)
\label{egn}
\end{eqnarray}
with $\mu,\lambda = 1,2,\cdots, 8$. For briefty, we here suppress the spin index on the wave function and energy. With the result of the wavefunctions, the order parameters can be determined self-consistently. The formula for the order parameters in terms of the wavefunctions are presented in Appendix A. 

For the current to be spin polarized, we have $\Delta_{\sigma}^{ll'} = \sigma\Delta^{ll'}$. To satisfy this condition, we need to set $\Delta^{a'b'}= - \Delta^{ab}$ and $\Delta^{aa'} = \Delta^{bb'}$. Then under the transforms $\sigma \to -\sigma$, $a \to a'$, $b \to b'$, and $p_y \to -p_y$, the effective Hamiltonian is unchanged. 

\section{Numerical results and discussion}

To numerically solve the model, we first need to determine the seven interactions $v^{ab}_{\mu\nu}(0)$ in the above equations for the order parameters. By considering the symmetry of the lattice, only three different values for the interactions are in question: $v^{aa}_{12}(0) = v^{bb}_{12}(0)$, $v^{ab}_{11}(0)=v^{a'b'}_{11}(0)$, and $v^{aa'}_{11}(0) = v^{ab'}_{11}(0) = v^{bb'}_{11}(0)$. In Appendix B, we present the formula for these interactions in terms of sums over the original lattice points and show the relation between $v^{aa}_{12}(0)$ and the interaction parameter in our previous work.\cite{Yan3}. In unit of $\epsilon_0 = 1$, the three interactions are determined as $v^{aa}_{12}(0)$ = 4.25, $v^{ab}_{11}(0)$ = 4.77, and  $v^{ab'}_{11}(0)$ = 2.94. With $v^{aa}_{12}(0)$ = 4.25 (so $v_s$ = 6.37 as given in previous work\cite{Yan3}), we get $\Delta_{\sigma}^a \approx$ 1 meV $\equiv \Delta_0$ (that is the experimental data\cite{Velasco}) by setting all the inter-sublattice current orderings to zero. 

\begin{figure}[b]
\centerline{\epsfig{file=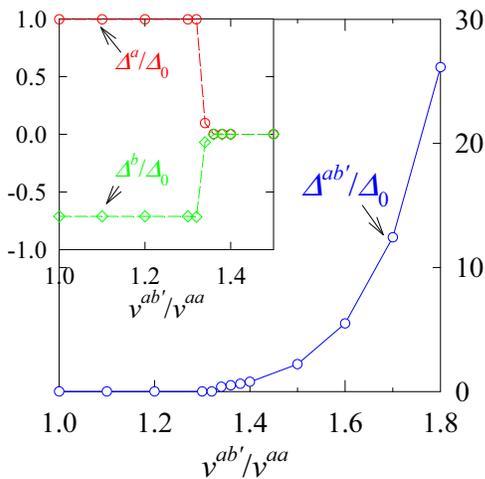,height=6.8 cm}}
\caption{(Color online) Current order $\Delta^{ab'}$ as function of the relative interaction strength $v^{ab'}/v^{aa}$ with $v^{ab}_{11}(0)/v^{aa}_{12}(0)$ = 5. Inset: orders $\Delta^{a}$ and $\Delta^{b}$.} 
\end{figure} 

We have solved the model at various relative strengths $v^{ab}_{11}(0)/v^{aa}_{12}(0)$ and $v^{ab'}_{11}(0)/v^{aa}_{12}(0)$ and at zero temperature. To a wide range of the relative strengths $v^{ab}_{11}(0)/v^{aa}_{12}(0)$ and $v^{ab'}_{11}(0)/v^{aa}_{12}(0)$, we find that the orderings $\Delta_{\sigma}^{ab}$, $\Delta_{\sigma}^{a'b'}$, $\Delta_{\sigma}^{aa'}$, and $\Delta_{\sigma}^{bb'}$ vanish. The interlayer current ordering $\Delta_{\sigma}^{ab'}$ appears when the interaction $v^{ab'}_{11}(0)$ is strong enough. Figure 6 shows the result for $\Delta^{ab'}\equiv\Delta_{\sigma}^{ab'}$ of spin-up ($\sigma = +$) electrons as a function of the strength $v^{ab'}/v^{aa} \equiv v^{ab'}_{11}(0)/v^{aa}_{12}(0)$ with $v^{ab}_{11}(0)/v^{aa}_{12}(0)$ = 5. (Since here $\Delta_{\sigma}^{ab}$, $\Delta_{\sigma}^{a'b'}$, $\Delta_{\sigma}^{aa'}$, and $\Delta_{\sigma}^{bb'}$ vanish, the current orderings $\Delta_{\sigma}^{a}$, $\Delta_{\sigma}^{b}$, and $\Delta_{\sigma}^{ab'}$ can be spin polarized.) It is seen that $\Delta^{ab'}$ shows up as increasing the strength $v^{ab'}/v^{aa}$ over 1.33, while the orders $\Delta^a\equiv\Delta_{+}^a$ and $\Delta^b\equiv\Delta_{+}^b$ as shown in the inset in Fig. 6 decrease sharply at this strength. The critical strength 1.33 is larger than the physical strength 0.69 determined by the above parameters. At $v^{ab'}/v^{aa} < 1.33$, only $\Delta^a$ and $\Delta^b$ are finite but all other orderings vanish. The result for $\Delta^a$ and $\Delta^b$ actually coincides our previous one.\cite{Yan3} From Fig. 6, we conclude that (1) the intra-sublattice current orderings cannot coexist with the inter-sublattice current orderings, and (2) within the physical strength of the interactions, only the former can appear.

We here discuss the effect of external electric field on the current orderings. Suppose an external electric field $E$ is applied perpendicularly to the BLG plane. Then, there is a potential difference $2u=Eed/\epsilon$ (with $\epsilon$ a screening constant) between the two layers. For small $u$, the external field does not affect the appearance of $\Delta^{ab'}$ since the critical interaction $v^{ab'}/v^{aa}$ = 1.33 is very strong. On the other hand, the intra-sublattice current orderings delicately depend on the external field. Figure 7 shows the intra-sublattice current orders $\Delta^{a}$ and $\Delta^{b}$ as functions of the external potential $u$. It is seen from Fig. 7 that the two order parameters decrease with increasing the potential and eventually vanish. 

\begin{figure}[t]
\centerline{\epsfig{file=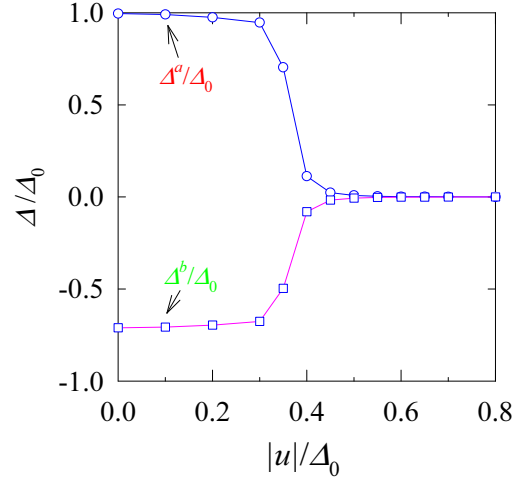,width=7 cm}}
\caption{(Color online) Current orders $\Delta^{a}$ and $\Delta^{b}$ as functions of the external potential $u$.} 
\end{figure} 

We have also studied the model dependence of the current orderings. The model is modified with including the so-called $\gamma_3$ hopping term between the $a$ and $b'$ atoms. Except the $t$ and $t_1$ hopping terms, the $\gamma_3$ term with $\gamma_3 \approx 3t_1/4$ is the strongest one among the remaining hoppings.\cite{Koshino} At low energy, this hopping gives rise to the elements $H^{eff}_{16} = H^{eff\dagger}_{61} = v_3(p_x+ip_y)$ ($K$ valley interlayer coupling) and $H^{eff}_{38} = H^{eff\dagger}_{83} = v_3(-p_x+ip_y)$ (-$K$ valley interlayer coupling) with $v_3=\sqrt{3}\gamma_3/2$ in the effective Hamiltonian. By including this hopping, the critical interaction strength for the appearance of $\Delta^{ab'}$ is pushed to higher value $v^{ab'}/v^{aa} \approx $ 1.5. At $v^{ab'}/v^{aa} > 1.5$, $\Delta^{ab'}$ is almost the same as that given in Fig. 6, $\Delta^a = 0$ and $\Delta^b =0$. While at $v^{ab'}/v^{aa} < 1.5$, $\Delta^{ab'} = 0$, $\Delta^{a}$ and $\Delta^{b}$ are two finite constants with reductions (due to the additional hopping on them) less than $2\%$. This result means the above conclusion obtained from Fig. 6 is unchanged.

We end this section with checking the approximation in Eq. (\ref{se1}) that the function Im$\langle c^{\dagger}_{kl'\nu\sigma}c_{kl\mu\sigma}\rangle$ is sizable at small $k$. To see it, we define the three possible nonvanishing functions
\begin{eqnarray}
f_a(p) &=& 2\sqrt{3}\int^{\pi}_{-\pi}\frac{d\theta}{2\pi} 
{\rm Im}\langle a^{\dagger}_{k1\sigma}a_{k2\sigma}\rangle  \nonumber\\
f_b(p) &=& 2\sqrt{3}\int^{\pi}_{-\pi}\frac{d\theta}{2\pi} 
{\rm Im}\langle b^{\dagger}_{k1\sigma}b_{k2\sigma}\rangle  \nonumber\\
f_{ab'}(p) &=& 6\int^{\pi}_{-\pi}\frac{d\theta}{2\pi} 
{\rm Im}\langle a^{\dagger}_{k1\sigma}b'_{k1\sigma}\rangle   \nonumber
\end{eqnarray}
where $\theta$ is the angle of the momentum $\vec p$. The behaviors of these functions are shown in Fig. 8. The left panel presents the functions $f_a(p)$ (main panel) and $f_b(p)$ (inset) at the interaction strength $v^{ab'}/v^{aa} = 1.$. The right panel is $f_{ab'}(k)$ at $v^{ab'}/v^{aa} = 1.5$. It is clear that these functions are nonvanishing at small $p$. This justifies our approximation in Eq. (\ref{se1}).

\begin{figure}[t]
\centerline{\epsfig{file=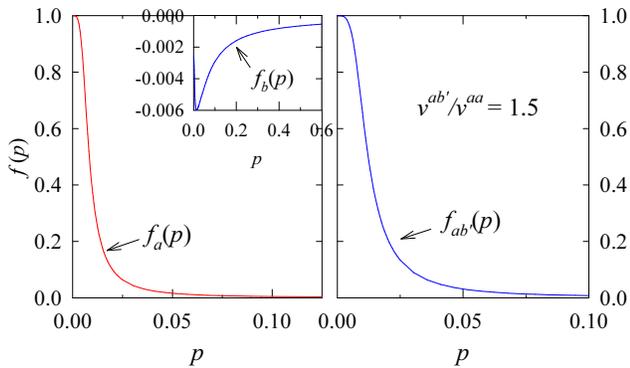,width=8.5 cm}}
\caption{(Color online) Left: Functions $f_a(p)$ (main panel) and $f_b(p)$ (inset) at $v^{ab'}/v^{aa} = 1$. Right: Function $f_{ab'}(p)$ at $v^{ab'}/v^{aa} = 1.5$.} 
\end{figure} 

\section{Summary} 

By taking into account all possible current orderings, we have derived an eight-band model for interacting electrons in BLG. We have solved the model to a wide range of the interaction strength and found that among the inter-sublattice currents only the interlayer current between $a$ and $b'$ atoms can appear when the interactions between electrons at these atoms are strong enough. The critical interaction strength for the current orderings between the $a$ and $b'$ sublattices is about 2 times stronger than the physical interaction strength. We have thus confirmed that within the range of physical interaction strength only the intra-sublattice current orderings are possible. Physically, because the inter-sublattice current orderings break the translational invariance, it is not likely to appear at weakly interacting systems; much stronger strength of interactions is needed for constructing a more symmetry broken state. On the other hand, since the intra-sublattice current orderings maintain the system's translational invariance, it can be realized at relatively weak interactions. The triangle lattices in the BLG are the favorite spaces for the current orderings. In such a space, the current orderings can survive in long range; since the currents flow along three dictions, the current density and the total current are zero satisfying the requirement of the equilibrium condition.

\acknowledgments

This work was supported by the National Basic Research 973 Program of China under Grants No. 2011CB932700 and No. 2012CB932302, and the Robert A. Welch Foundation under Grant No. E-1146.

\begin{center}
{\bf APPENDIX A: Order parameters}
\end{center}
\renewcommand{\theequation}{A\arabic{equation}}
\setcounter{equation}{0}

Here, we express the formula for calculating the order parameters from the wavefunctions. Denote the eigen operator as $\Phi_{\lambda}(p)$. The original operator $\psi_{k\sigma}$ can be expressed as 
\begin{eqnarray}
\psi_{k\sigma}=(I_2\otimes T^r)\phi(p)\Phi(p)       \label{trsf}
\end{eqnarray}
where $T^r$ is a 6$\times$4 matrix obtained by deleting the last two columns in $T$, $\phi(p)$ is an 8$\times$8 matrix with elements $\phi_{\mu\lambda}(p)$, and $\Phi(p)$ is an 8-component vector with components $\Phi_{\lambda}(p)$. We here delete the last two columns in $T$ because they are related in the transformation to the eigen-states of matrix $D$ given by Eq. (\ref{dm}) that has been dropped away in obtaining the eight-band model given by Eq. (\ref{heff}). Explicitly, the matrix is
\begin{eqnarray}
T^r=\frac{1}{\sqrt{3}}\begin{pmatrix}
e^{i\alpha}& 1\\
e^{-i\alpha}&e^{-i\alpha}\\
1& e^{i\alpha}\\
\end{pmatrix}\otimes\sigma_0. \nonumber
\end{eqnarray}
With the transform given by Eq. (\ref{trsf}), the order parameters are then calculated as
\begin{eqnarray}
\Delta_{\sigma}^a &=&\frac{v^{aa}_{12}(0)}{2N}\sum_{k\lambda}[\phi^{\dagger}_{1\lambda}(p)\phi_{1\lambda}(p)
-\phi^{\dagger}_{3\lambda}(p)\phi_{3\lambda}(p)]f_{\lambda}(p), \nonumber\\
\Delta_{\sigma}^b &=&\frac{v^{bb}_{12}(0)}{2N}\sum_{k\lambda}[\phi^{\dagger}_{2\lambda}(p)\phi_{2\lambda}(p)
-\phi^{\dagger}_{4\lambda}(p)\phi_{4\lambda}(p)]f_{\lambda}(p), 
\nonumber\\
\Delta_{\sigma}^{ab} &=&\frac{v^{ab}_{11}(0)}{N}\sum_{k\lambda}{\rm Im}[\phi^{\dagger}_{1\lambda}(p)\phi_{2\lambda}(p)
+\phi^{\dagger}_{3\lambda}(p)\phi_{4\lambda}(p) \nonumber\\
&&~~+e^{-i\alpha}\phi^{\dagger}_{1\lambda}(p)\phi_{4\lambda}(p)
+e^{i\alpha}\phi^{\dagger}_{3\lambda}(p)\phi_{2\lambda}(p)]f_{\lambda}(p), \nonumber\\
\Delta_{\sigma}^{a'b'} &=&\frac{v^{a'b'}_{11}(0)}{N}\sum_{k\lambda}{\rm Im}[\phi^{\dagger}_{5\lambda}(p)\phi_{6\lambda}(p)
+\phi^{\dagger}_{7\lambda}(p)\phi_{8\lambda}(p) \nonumber\\
&&~~+e^{-i\alpha}\phi^{\dagger}_{5\lambda}(p)\phi_{8\lambda}(p)
+e^{i\alpha}\phi^{\dagger}_{7\lambda}(p)\phi_{6\lambda}(p)]f_{\lambda}(p), 
\nonumber\\
\Delta_{\sigma}^{aa'} &=&\frac{v^{aa'}_{11}(0)}{N}\sum_{k\lambda}{\rm Im}[\phi^{\dagger}_{1\lambda}(p)\phi_{5\lambda}(p)
+\phi^{\dagger}_{3\lambda}(p)\phi_{7\lambda}(p)\nonumber\\
&&~~+e^{-i\alpha}\phi^{\dagger}_{1\lambda}(p)\phi_{7\lambda}(p)
+e^{i\alpha}\phi^{\dagger}_{3\lambda}(p)\phi_{5\lambda}(p)]f_{\lambda}(p), \nonumber\\
\Delta_{\sigma}^{bb'} &=&\frac{v^{bb'}_{11}(0)}{N}\sum_{k\lambda}{\rm Im}[\phi^{\dagger}_{2\lambda}(p)\phi_{6\lambda}(p)
+\phi^{\dagger}_{4\lambda}(p)\phi_{8\lambda}(p) \nonumber\\
&&~~+e^{-i\alpha}\phi^{\dagger}_{2\lambda}(p)\phi_{8\lambda}(p)
+e^{i\alpha}\phi^{\dagger}_{4\lambda}(p)\phi_{6\lambda}(p)]f_{\lambda}(p), 
\nonumber\\
\Delta_{\sigma}^{ab'} &=&\frac{v^{ab'}_{11}(0)}{N}\sum_{k\lambda}{\rm Im}[\phi^{\dagger}_{1\lambda}(p)\phi_{6\lambda}(p)
+\phi^{\dagger}_{3\lambda}(p)\phi_{8\lambda}(p) \nonumber\\
&&~~+e^{-i\alpha}\phi^{\dagger}_{1\lambda}(p)\phi_{8\lambda}(p)
+e^{i\alpha}\phi^{\dagger}_{3\lambda}(p)\phi_{6\lambda}(p)]f_{\lambda}(p), \nonumber
\end{eqnarray}
where $f_{\lambda}(p) = f[E_{\lambda}(p)] = \langle\Phi^{\dagger}_{\lambda}(p)\Phi_{\lambda}(p)\rangle$ is the Fermi distribution function. Here, the $k$-summation can be replaced with the $p$-integration according to
\begin{eqnarray}
\frac{1}{N} \sum_{k}= \int \frac{d\vec k}{(2\pi)^2} = \frac{3\sqrt{3}}{2}\int\frac{d\vec p}{(2\pi)^2} \nonumber
\end{eqnarray}
with a cutoff $p_c = 1$ for the largest momentum. We thus have obtained the self-consistent equations for determining the order parameters.

\begin{center}
{\bf APPENDIX B: Interaction parameters}
\end{center}
\renewcommand{\theequation}{B\arabic{equation}}
\setcounter{equation}{0}

We here derive the formula for these interactions so that they are given by sums over the original lattice points instead of over the superlattice ones. By so doing, we will clearly see how the present model reduces to the previous one. Denoting $\vec r^{ll'}_{\mu\nu} = \vec \rho^{ll'}_{\mu\nu} + \vec z$ with $\vec \rho^{ll'}_{\mu\nu}$ the planar component and $z = 0$ or $d$ in $\hat z$ direction respectively for intralayer or interlayer interactions, we expand $v(\vec R + \vec r^{ll'}_{\mu\nu})$ as
\begin{widetext}
\begin{eqnarray}
v(|\vec R + \vec r^{ll'}_{\mu\nu}|) &=& \frac{1}{V} \sum_{\vec q}v(q,z)\exp[i\vec q\cdot(\vec R+\vec\rho^{ll'}_{\mu\nu})] \nonumber\\
&=& \frac{1}{V} {\sum_{\vec q,n}}'v(|\vec Q_n+\vec q|,z)\exp[i(\vec Q_n+\vec q)\cdot\vec\rho^{ll'}_{\mu\nu}]\exp[i\vec q\cdot\vec R] \label {vn}
\end{eqnarray}
where $V$ is the area of the graphene layer, $\vec R$ represents the position of an unit cell in the superlattice, $\vec q$-summation in the second line runs over the first BZ given by the smaller hexagon in Fig. 3, and $n$ over all the points including the red-circles and the blue points in Fig. 3. Here, we have used the relation $\exp(i\vec Q_n\cdot\vec R) = 1$. From the definition for $v^{ll'}_{\mu\nu}(q)$ and Eq. (\ref{vn}), we get
\begin{eqnarray}
v^{ll'}_{\mu\nu}(q) &\equiv& \sum_{\vec R}v(|\vec R + \vec r^{ll'}_{\mu\nu}|)\exp(-i\vec q\cdot\vec R) \nonumber\\
&=& \frac{1}{3v_0} 
\sum_{n}v(|\vec Q_n+\vec q|,z)\exp[i(\vec Q_n+\vec q)\cdot\vec\rho^{ll'}_{\mu\nu}] \nonumber
\end{eqnarray}
with $v_0 = \sqrt{3}/2$ as the area of the unit cell of the original honeycomb lattice. For $v^{ll'}_{\mu\nu}(0)$, we have
\begin{eqnarray}
v^{ll'}_{\mu\nu}(0) &=& \frac{1}{3v_0} 
{\sum_m}'[v(Q_m,z)+v(|\vec Q_m+\vec K|,z)\exp(i\vec K\cdot\vec\rho^{ll'}_{\mu\nu})+ v(|\vec Q_m-\vec K|,z)\exp(-i\vec K\cdot\vec\rho^{ll'}_{\mu\nu})]\exp(i\vec Q_m\cdot\vec\rho^{ll'}_{\mu\nu}), \label{vmn}
\end{eqnarray}
where $m$ sums over the blue points in Fig. 3. Here, noting that $\exp(i\vec Q_m\cdot\vec\rho^{ll}_{\mu\nu}) = 1$ and $\cos(\vec K\cdot\vec\rho^{ll}_{\mu\nu}) = -1/2$ (for $\rho^{ll}_{\mu\nu} = \pm 1$) for the atoms in the same sublattice, we have 
\begin{eqnarray}
v^{ll}_{\mu\nu}(0) &=& \frac{1}{3v_0} 
{\sum_m}'[v(Q_m,z)-v(|\vec Q_m+\vec K|,z)], ~~{\rm for}~\vec\rho^{ll}_{\mu\nu}= \pm 1. \label{vmaa}
\end{eqnarray}
As stated after Eq. (\ref{sab}) in Sec. II, we consider the vectors $\vec\rho^{ll'}_{\mu\nu}$'s between the atoms in the small unit cell containing only four atoms $a_1$, $b_1$, $a'$ (beneath $b_1$) and $b_1'$. Then, for $\vec r$ being a position of the unit cell containing the four atoms, we have the expansion
\begin{eqnarray}
v(|\vec r + \vec r^{ll'}_{\mu\nu}|) 
&=& \frac{1}{V} {\sum_{\vec q,m}}'v(|\vec Q_m+\vec q|,z)\exp[i(\vec Q_m+\vec q)\cdot\vec\rho^{ll'}_{\mu\nu}+i\vec q\cdot\vec r], ~~ {\rm for}~ l \ne l' \nonumber\\
v(r) &=& \frac{1}{V} {\sum_{\vec q,m}}'v(|\vec Q_m+\vec q|,z)\exp(i\vec q\cdot\vec r), ~~ {\rm for} ~l = l'\nonumber
\end{eqnarray}
where the $\vec q$-summation runs over the first BZ given by the larger hexagon in Fig. 3. By definition, we have the Fourier components
\begin{eqnarray}
\overline v^{ll'}_{\mu\nu}(q) &\equiv& \sum_{\vec r}v(|\vec r + \vec r^{ll'}_{\mu\nu}|)\exp(-i\vec q\cdot\vec r) \nonumber\\
&=& \frac{1}{v_0} 
\sum_{m}v(|\vec Q_m+\vec q|,z)\exp[i(\vec Q_m+\vec q)\cdot\vec\rho^{ll'}_{\mu\nu}], ~~~~{\rm for}~ l\ne l' \label{vm1}\\
\overline v^{ll}_{\mu\nu}(q) &\equiv& \sum_{\vec r}v(r)\exp(-i\vec q\cdot\vec r) 
= \frac{1}{v_0} 
\sum_{m}v(|\vec Q_m+\vec q|,z). \label{vm2}
\end{eqnarray}
\end{widetext}
By applying Eq. (\ref{vm1}) in Eq. (\ref{vmn}), we get for $l \ne l'$
\begin{eqnarray}
v^{ll'}_{\mu\nu}(0) &=& \frac{1}{3} 
[\overline v^{ll'}_{\mu\nu}(0) +\overline v^{ll'}_{\mu\nu}(\vec K)+\overline v^{ll'}_{\mu\nu}(-\vec K)]\nonumber\\
&=& \frac{1}{3}\sum_{\vec r}v(|\vec r + \vec r^{ll'}_{\mu\nu}|)[1+2\cos(\vec K\cdot\vec r)]. 
\label{vf}
\end{eqnarray}
From Eqs. (\ref{vmaa}) and (\ref{vm2}), we have
\begin{eqnarray}
v^{ll}_{12}(0) &=& \frac{1}{3} 
[\overline v^{ll}_{12}(0) -\overline v^{ll}_{12}(\vec K)]\nonumber\\
&=& \frac{1}{3}\sum_{\vec r}v(r)[1-\cos(\vec K\cdot\vec r)] \nonumber\\
&=& \frac{2}{3}\sum_{\vec r}v(r)\sin^2(\vec K\cdot\vec r) 
\label{vaaf}
\end{eqnarray}
where in the last line we have used the fact $\cos(\vec K\cdot\vec r) = \cos(2\vec K\cdot\vec r)$ since the difference between $2\vec K$ and $-\vec K$ is a lattice vector (blue) in the reciprocal space. Using the interaction parameter $v_s$ as defined in our previous work,\cite{Yan3} we have $v^{ll}_{12}(0) = 2v_s/3$. 
Applying this result in the expressions for $\Delta_{\sigma}^a$ and $\Delta_{\sigma}^b$ in Appendix A and recognizing $\sum_{\lambda}f_{\lambda}(p)[\phi^{\dagger}_{1(2)\lambda}(p)\phi_{1(2)\lambda}(p)-\phi^{\dagger}_{3(4)\lambda}(p)\phi_{3(4)\lambda}(p)]$ as the valley difference of the electron distribution function in $a$ ($b$) sublattice, we reach the same forms for $\Delta_{\sigma}^a$ and $\Delta_{\sigma}^b$ as in our previous work.\cite{Yan3}


\begin{thebibliography}{99}

\bibitem{Ohta} T. Ohta, A. Bostwick, T. Seyller, K. Horn, E. Rotenberg, Science {\bf 313}, 951 (2006).

\bibitem{Oostinga} J. B. Oostinga, H. B. Heersche, X. Liu, A. F. Morpurgo, and L. M. K. Vanderspen, Nature Mater. {\bf 7}, 151 (2008).

\bibitem{McCann} E. McCann, Phys. Rev. B {\bf 74}, 161403(R) (2006).

\bibitem{Castro} E. V. Castro, K. S. Novoselov, S. V. Morozov, N. M. R. Peres, J. M. B. L. dos Santos, J. Nilsson, F. Guinea, A. K. Geim, and A. H. Castro Neto, Phys. Rev. Lett. {\bf 99}, 216802 (2007).

\bibitem{Weitz} R. T. Weitz, M. T. Allen, B. E. Feldman, J. Martin, and A. Yacoby, Science {\bf 330}, 812 (2010).

\bibitem{Freitag} F. Freitag, J. Trbovic, M. Weiss, and C. Sch\"onenberger, Phys. Rev. Lett. {\bf 108}, 076602 (2012).

\bibitem{Velasco} J. Velasco Jr., L. Jing, W. Bao, Y. Lee, P. Kratz, V. Aji, M. Bockrath, C. N. Lau, C. Varma, R. Stillwell, D. Smirnov, F. Zhang, J. Jung, and A. H. MacDonald, Nature Nanotech. {\bf 7}, 156 (2012).

\bibitem{Bao} W. Bao, J. Velasco Jr., L. Jing, F. Zhang, B. Standley, D. Smirnov, M. Bockrath A. H. MacDonald, and C. N. Lau, Proc. Natl. Acad. Sci. USA {\bf 109}, 10802 (2012).

\bibitem{Elferen} H. J. van Elferen, A. Veligura, E. V. Kurganova, U. Zeitler, J. C.
Maan, N. Tombros, I. J. Vera-Marun, and B. J. van Wees, Phys.
Rev. B {\bf 85}, 115408 (2012).

\bibitem{Velasco1} J. Velasco, Jr., Y. Lee, Z. Zhao, L. Jing, P. Kratz, M. Bockrath,
and C. N. Lau, Nano Lett. {\bf 14}, 1324 (2014). 

\bibitem{Min} H. K. Min, G. Borghi, M. Polini, and A. H. MacDonald, Phys. Rev. B {\bf 77}, 041407(R) (2008); F. Zhang, H. K. Min, M. Polini, and A. H. MacDonald, Phys. Rev. B {\bf 81}, 041402(R) (2010);
A. H. MacDonald, J. Jung, and F. Zhang, Phys. Scr. {\bf T146}, 014012 (2012).

\bibitem{Nandkishore} R. Nandkishore and L. Levitov, Phys. Rev. Lett. {\bf 104}, 156803 (2010).

\bibitem{Zhang2} F. Zhang and A. H. MacDonald, Phys. Rev. Lett. {\bf 108}, 186804 (2012).

\bibitem{Nilsson} J. Nilsson, A. H. Castro Neto, N. M. R. Peres, and F. Guinea, Phys. Rev. B {\bf 73}, 214418 (2006).

\bibitem{Gorbar} E. V. Gorbar, V. P. Gusynin, V. A. Miransky, and I. A. Shovkovy, Phys. Rev. B {\bf 85}, 235460 (2012).

\bibitem{Jung} J. Jung, F. Zhang, and A. H. MacDonald, Phys. Rev. B {\bf 83}, 115408 (2011).

\bibitem{Nandkishore1} R. Nandkishore and L. Levitov, Phys. Rev. B {\bf 82}, 115124 (2010).

\bibitem{Zhang1} F. Zhang, J. Jung, G. A. Fiete, Q. Niu, and A. H. MacDonald, Phys. Rev. Lett. {\bf 106}, 156801 (2011).

\bibitem{Milovanovic} M. V. Milovanovi\'c and S. Predin, Phys. Rev. B 86, 195113 (2012).

\bibitem{Zhu} L. J. Zhu, V. Aji, and C. M. Varma, Phys. Rev. B {\bf 87}, 035427 (2013).

\bibitem{Yan1} X.-Z. Yan and C. S. Ting, Phys. Rev. B {\bf 86}, 235126 (2012). 

\bibitem{Yan2} X.-Z. Yan and C. S. Ting, Phys. Rev. B {\bf 88}, 045410 (2013). 

\bibitem{Yan3} X.-Z. Yan and C. S. Ting, Phys. Rev. B {\bf 89}, 201108(R) (2014). 

\bibitem{san-jose} P. San-Jose, R. V. Gorbachev, A. K. Geim, K. S. Novoselov, and F. Guinea, Nano. Lett. {\bf 14}, 2052 (2014).

\bibitem{Lemonik} Y. Lemonik, I. L. Aleiner, C. Toke, and V. I. Fal'ko, Phys. Rev. B {\bf 82}, 201408(R) (2010); O. Vafek and K. Yang, Phys. Rev. B {\bf 81}, 041401(R) (2010).

\bibitem{Tatar} R. C. Tatar and S. Rabii, Phys. Rev. B {\bf 25}, 4126 (1982).

\bibitem{LMZ} L. M. Zhang, Z. Q. Li, D. N. Basov, M. M. Fogler, Z. Hao and M. C. Martin, Phys. Rev. B {\bf 78}, 235408 (2008).

\bibitem{Varma} C. M. Varma, Phys. Rev. Lett. {\bf 83}, 3538 (1999).

\bibitem{Hwang} E. H. Hwang and S. Das Sarma, Phys. Rev. Lett. {\bf 101}, 156802 (2008).

\bibitem{Koshino} M.Koshino and E. McCann, Phys. Rev. B {\bf 83}, 165443 (2011).

\end{thebibliography}
\end{document}